%% file: main.tex
\documentclass[sigconf]{acmart}
\usepackage{adjustbox}
\usepackage{mdframed}
\usepackage{tabularx}
\usepackage{multirow}
\usepackage{listings}
\usepackage{tcolorbox}
\usepackage{ragged2e} 
\usepackage{subfig}
\usepackage[dvipsnames]{xcolor}

\newcommand{\rev}[1]{\textcolor{Black}{#1}}
\newcolumntype{Y}{>{\Centering\arraybackslash}X}

\acmYear{2026}\copyrightyear{2026}
\setcopyright{cc}
\setcctype[4.0]{by}

\begin{document}

\title{Agentic Code Review in the Terminal: A Trajectory-Level Analysis of Behavior, Cost, and Human-Alignment}

\author{Wachiraphan Charoenwet}
\affiliation{%
  \institution{The University of Melbourne}
  \country{Australia.}
}
\author{Kla Tantithamthavorn}
\affiliation{%
  \institution{Monash University}
  \country{Australia.}
}

\author{Patanamon Thongtanunam}
\affiliation{%
  \institution{The University of Melbourne}
  \country{Australia.}
}

\author{Hong Yi Lin}
\affiliation{%
  \institution{The University of Melbourne}
  \country{Australia.}
}

\author{Minwoo Jeong}
\affiliation{%
  \institution{Atlassian}
  \country{USA.}
}

\author{Ming Wu}
\affiliation{%
  \institution{Atlassian}
  \country{USA.}
}

\begin{abstract}

Agentic code review in terminal-based environments enables early feedback during local development before pull request creation. 
However, existing evaluations remain performance-centric and fail to capture the dynamic behaviors of repository-grounded agentic reviewers. 
Understanding these behaviors is critical for identifying how agentic reviewers succeed, fail, and incur hidden operational costs in practice. 
Then, we analyze the reviewers' behavior based on their trajectories.
Our results show that agentic reviewers achieve higher review precision, but incur substantial exploration and validation overhead, while successful reviews are associated with stronger planning and less downstream validation. 
These findings highlight the potential benefits of trajectory-aware and cost-sensitive evaluation of future agentic code review systems.


\end{abstract}

\maketitle

\section{Introduction}
Agentic code review in the terminal moves review earlier in the software lifecycle, enabling developers to obtain feedback during local development before opening a pull request. 
Unlike static LLM-based review approaches that operate on fixed prompts based on the information within a Pull Request (PR), terminal-based agentic code reviewers can explore the surrounding context within the repository.
Broadly speaking, agents can autonomously decide \emph{what} context in the repository to retrieve, \emph{where} to look next, and \emph{which} tools to invoke to review the code.
When effective, this repository-grounded approach can improve review relevance and precision; however, when ineffective, it incurs substantial overhead through inefficient navigation and unnecessary context retrieval.
Existing code review benchmarks are not designed for this setting~\cite{hu2025contextcrbench,zhang2026aacrbench}. 
Most assume static evaluation and assess review quality solely from the final comments generated. 
This is mismatched to agentic code review, where performance depends not only on outputs but also on repository-grounded behavior where the agent plans, navigates the repository, reasons over, and validates its generation against the relevant context~\cite{Liu2026Graphectory, ou-etal-2025-agentdiagnose, kim2026trajevaldecomposingcodeagent}. 
These intermediate actions and reasoning steps, as well as underlying processes and tool uses, are typically recorded as an \textbf{\textit{agent trajectory}}, which
remain largely unexamined in prior work. 
As a result, current evaluations provide limited insight into whether repository context improves code review performance, how agentic reviewers allocate effort, and what costs arise from multi-step review behavior.

In this paper, we investigate repository-grounded agentic code review in the terminal through the lens of agent behavior (i.e., trajectories), analyzing tool use, operational costs, and human alignment.
To support this investigation, we reconstruct a repo-level, human-verified code review dataset in a terminal-based environment. 
Using this setup, we evaluate four state-of-the-art terminal-based agentic reviewers, Claude Code, Gemini CLI, Rovo Dev, and Code Rabbit, against static LLM reviewers. 
Our results show that agentic reviewers improve review precision through repository-grounded exploration, but incur substantial exploration and validation overhead. 
Trajectory analysis further reveals that successful reviews are associated with stronger planning, while excessive validation often correlates with failure. 
Together, these trajectory-centric findings complement outcome-level benchmark results by revealing how reviewers arrive at review decisions and exposing trade-offs among exploration, deliberation, efficiency, and review quality.

\noindent\textbf{Contributions.} This paper makes the following contributions:

\begin{itemize}
\setlength{\leftskip}{-10pt} 
    \item AgenticCR-Verified, a human-verified real-world code review dataset, reconstructed for agentic code review.
    \item An empirical evaluation of agentic code review in terminal and static LLM reviewers based on agent behavior (trajectories), operational costs, and human alignment.
    \item An association between agentic trajectories and review success based on human alignment.
\end{itemize}

\section{Background and Related Work}
\label{sec:terminal-based-agent}

\textbf{Agentic Code Reviews in a Nutshell.} Agentic code review refers to LLM-driven code review in which an LLM agent autonomously retrieves repository context, invokes tools, and refines judgments before producing comments. 
Such agents can operate in IDEs, pull request interfaces, or command-line terminals. 
In this work, we focus on \emph{terminal-based} agentic code review, where reviewers operate directly within a local repository via a command-line interface (CLI), using commands such as \texttt{grep}, \texttt{find}, and \texttt{cat} to inspect code and gather evidence. 
Unlike static LLMs that generate a comment based on fixed inputs, terminal-based agentic reviewers follow an iterative observe--act loop, interleaving reasoning with tool use to retrieve repository context dynamically~\cite{yao2023react, Schick2023Toolformer, merrill2026terminal, lin2026cligym}. 
This repository-grounded behavior helps explore and validate relevant evidence spanning across files, abstractions, and dependencies.

\begin{figure}[t]
    \centering
    \includegraphics[width=\linewidth]{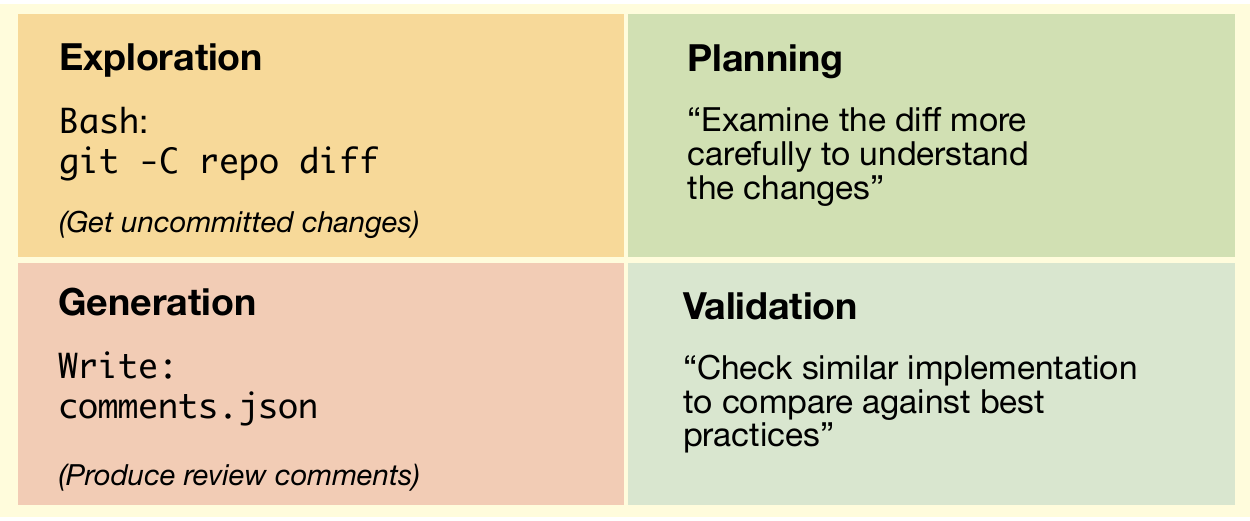}
\caption{Representative examples of trajectory code review phase. Messages and tool calls are abbreviated for readability.}
    \label{fig:phase-example}
    \vspace{-5mm}
\end{figure}


\label{sec:agentic_trajectory}
\textbf{Agentic trajectory.} A trajectory is the sequence of intermediate actions and reasoning steps an agent performs while completing a review task~\cite{Liu2026Graphectory, ou-etal-2025-agentdiagnose, kim2026trajevaldecomposingcodeagent}. 
Typically,  a trajectory is composed of five phases: (1) initialization (task setup), (2) exploration (retrieving repository context), (3) planning (intermediate deliberation and decision-making steps), (4) generation (producing review comments), and (5) validation (verifying or refining findings). 
Figure~\ref{fig:phase-example} shows examples of the agent action in each phase recorded in a trajectory. 
This phase-level abstraction outlines how agents allocate actions and reasoning across a review task.

\textbf{Code Review Automation Benchmark.} 
Prior code review benchmarks typically assume static, direct prompting settings, where LLMs generate review comments from fixed inputs such as code diffs and limited pull-request context~\cite{hu2025contextcrbench}.
These approaches rely on single-pass prompting without additional repository exploration or iterative evidence gathering~\cite{Adalsteinsson2025Rethinking, zhang2026sphinx}.
As a result, evaluation primarily focuses on syntactic human alignment, i.e., the similarity between human-written and LLM-generated comments~\cite{hu2025contextcrbench}, and operational cost is based on input and output token usage~\cite{merrill2026terminal}, providing limited insight into the underlying agent behaviors and reasoning processes.
Evaluating agent behaviors alongside outcomes deepens understanding of what makes reviews effective.

\section{Toward Evaluating Agentic Code Review}

In this section, we present our evaluation of agentic code review.

\subsection{Dataset Construction}


To construct our dataset, we began with~\citet{hu2025contextcrbench}'s dataset, which contains 68K code review discussions from 99 GitHub repositories across 10 programming languages. 
Each discussion is associated with a GitHub issue, pull request (PR), and diff hunks. 
Since the dataset of ~\citet{hu2025contextcrbench} is for evaluating static LLM code review, we reconstruct the dataset to suit the agentic code reviewer, which requires repository-level context by defining a \textit{code review task} as:
\[
T(repo, \textit{commit hash}, \textit{diff hunks}, issue, PR \rightarrow \textit{comments})
\]
where \textit{repo} is the GitHub repository, \textit{commit hash} identifies the reviewed commit, \textit{diff hunks} denote code changes under review, and \textit{issue} and \textit{PR} provide contextual information from the corresponding GitHub issue and pull request. The \textit{comments} are developer-written review discussions on the diff hunks, used as ground truth.


Across 68K review discussions, we identify 30K unique code review task candidates, i.e., unique triplets of \textit{(issue, PR, and commit)} and apply a three-stage filter. 
We first exclude repositories larger than 1GB for feasibility, retaining 90 repositories. 
\rev{Then, we retain only tasks whose commit IDs remain resolvable in the repository's commit tree and whose diff hunks match the reviewed changes via token matching; most excluded tasks have commits that no longer exist or contain mismatched changes, reducing the dataset to 1,078 tasks.}
Finally, we stratify tasks by the size of diff hunk into Small–Medium ($\leq$100 additions), Large (101--500), and Extra Large ($>$500) to capture variation in review complexity~\cite{Rigby2014PeerProjects} and ensure balanced sampling, yielding 421 tasks.





To reduce noise in the ground truth~\cite{Liu2025TooNoisyToLearn}, we manually verify review comments and discard tasks with bot-generated or non-human comments, non-substantive remarks (e.g., acknowledgments or greetings), high-level discussions not tied to specific changes, or extreme commits (>2,000 LOC) where meaningful review is not realistic.
After verification, the final dataset consists of 362 terminal-based code review tasks, each with a terminal-based environment, and the aligned diff hunks and developers' comments.
This structure enables both evaluation and trajectory analysis of agentic reviewers under a realistic repository-level context.

\subsection{Experimental Design}

\subsubsection{Evaluation Harness}
To ensure reproducible evaluation, we use \textsc{Harbor} as a unified execution harness~\cite{harbor2026framework}. 
Harbor provides a sandboxed terminal environment where an agentic reviewer executes a task and interacts with a repository.
This setup is essential as agentic review involves repository exploration and environment interaction. 
Running all agents within the same harness standardizes execution conditions and reduces confounding variations.

\subsubsection{Trajectory Representation}
To understand how agentic reviewers perform repository-grounded code review, we analyze their trajectories in addition to their final outputs. 
We represent trajectories using the ATIF (Agent Trajectory Interchange Format),\footnote{https://github.com/harbor-framework/harbor/blob/main/rfcs/0001-trajectory-format.md} a JSON-based schema that records step-level interactions, including actions, intermediate signals (when available), and environment responses, enabling consistent analysis across heterogeneous agents.

\subsubsection{Phase Annotation.}
\rev{We identify the phases of an action or reasoning step by matching keyword lists against each step's message text, reasoning content, and tool call descriptions (e.g., \textit{read}, \textit{grep}, or \textit{search} map to exploration; \textit{write} or \textit{generate} to generation; \textit{verify} or \textit{check} to validation; \textit{plan} or \textit{next} to planning). }
Given a trajectory, we identify the five common phases as described in Section~\ref{sec:agentic_trajectory}.
Exploration covers repository inspection, planning captures intent and sequencing, generation produces reviews, and validation involves verification and refinement.
System messages are excluded.

\subsubsection{Trajectory Metrics.}
We quantify agentic behavior using four phase-level signals: steps, tool calls, tokens, and cost. Specifically, we measure (1) steps per phase, capturing effort allocation; (2) tool calls per phase, indicating environment interaction (e.g., file access, search, execution); (3) input and output tokens per phase; and (4) estimated total cost based on token usage. 
This decomposition reveals how agents distribute effort and computational cost.


\begin{figure*}[t]
    \centering
    \includegraphics[width=\linewidth]{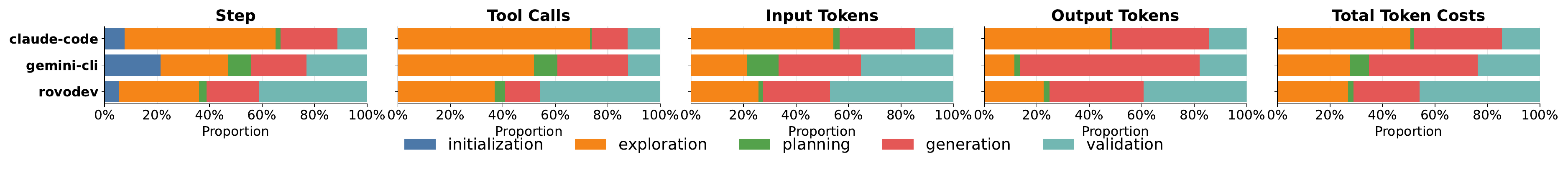}
    \caption{Phase-wise distribution of trajectory effort (mean across tasks). Rows denote agentic reviewers, subplots denote metrics (steps, tool calls, input and output tokens, cost). Segments indicate phase proportions.}
    \label{fig:trajectory-phase}
    \vspace{-5mm}
\end{figure*}

\subsubsection{Code Review Evaluation Metrics}
We evaluate agentic and static LLM reviewers at comment and task levels to support fine-grained review assessment and trajectory-level analysis.

\textbf{Comment-Level Metrics.}
To compare agentic and static LLM reviewers, we adopt two metrics: \textit{localization}, \rev{whether a generated comment falls within $\pm$5 lines of a human comment}, and \textit{human alignment}, whether it identifies a relevant issue, evaluated using an LLM-as-a-Judge framework~\cite{tantithamthavorn2026hallujudgere}.

\textbf{Code Review-Level Outcomes.}
\label{sec:trajectory_metric}
For trajectory analysis, we aggregate comment-level results into task level. 
A task is \emph{accomplished} if at least one generated comment satisfies both localization and human-alignment criteria. 
A task is \emph{partially accomplished} if comments satisfy only one of the two criteria. 
A task is \emph{unaccomplished} if none of the generated comments satisfy either criterion. 

\subsubsection{Studied Code Reviewers}

We study the following agentic code reviewers and static LLM reviewers.

\begin{itemize}\setlength{\leftskip}{-10pt} 
\item \textbf{Agentic Code Reviewers}: \textbf{Gemini CLI}, \textbf{Claude Code}, \textbf{Rovo Dev}, and \textbf{CodeRabbit} (specialized code review agent), all operating in terminal-based environments. 
\rev{Their main backend models are Gemini 2.5 Flash, Claude Sonnet 4.6 (for both Claude Code and Rovo Dev), and undisclosed model (for CodeRabbit).}
\item \textbf{Static LLM Reviewers}: agents that generate comments from fixed inputs (e.g., diffs and PR) without repository interactions.
For the LLM models, we use GPT-5.4, GPT-5.2, Claude Sonnet 4.6, Claude Sonnet 4.5, and Gemini 2.5 Pro.

\end{itemize}

\rev{These candidates are complementary: static LLM reviewers serve as a reference point for review without repository exploration, while agentic reviewers—the primary focus of this paper—represent production-grade, repository-aware agentic code review.}


\subsubsection{Code Review Task Instructions.}
To enable fair comparison, we standardize the review objective, directing agents to analyze code changes in the repository context and identify concrete and actionable issues.
For agentic reviewers, we encode Harbor's task instruction (\texttt{instruction.md}) through each reviewer's native CLI, standardizing both the environment and review objective across agents.
These controls allow us to compare agentic code reviewers and preserve their native interaction models. 
For static LLM reviewers, we use a similar prompt and experimental setup as~\citet{hu2025contextcrbench}'s work.
\rev{To reduce non-determinism, we set temperature to 0 for all reviewers.}


\section{Experimental Results}
We compare agentic reviewers against static LLM reviewers, then analyze agentic trajectories to understand underlying behaviors.

\input{tables/overall-performance}

\textbf{Overall Review Performance.}
Table~\ref{tab:overall-performance} reports the percentage of comments with correct localization, human alignment, and overall performance (both criteria satisfied).
Agentic reviewers achieve the strongest results, with \textbf{Rovo Dev} and \textbf{Claude Code} reaching 5.81\% and 5.60\%, respectively, driven primarily by improved localization (>20\% vs.\ 7\%--9\% for other reviewers).
Notably, this improvement occurs despite generating fewer comments, suggesting repository-grounded interaction improves comment quality rather than volume. 
\rev{The low absolute scores reflect the strictness of requiring both criteria simultaneously; agents may also identify valid issues not captured in the benchmark ground truth, which would appear as failures under our metric.}
We next investigate how agentic reviewers achieve these results through their trajectories, excluding CodeRabbit as it does not expose internal trajectories.


\textbf{RQ1: How do agents allocate effort across review trajectory phases?}
Agentic code reviewers exhibit distinct trajectory strategies, with effort dominated by exploration and validation, while planning remains minimal. 
Figure~\ref{fig:trajectory-phase} shows that in our repository-grounded setting, where agentic reviewers iteratively navigate and retrieve context from local codebases, effort distribution varies substantially. 
\textbf{Rovo Dev} is strongly validation-driven (41\% of steps, 46\% of tool calls, 46\% of cost), reflecting repeated verification and refinement after generation. 
In contrast, \textbf{Claude Code} is exploration-centric, allocating most effort to repository traversal (58\% of steps, 73\% of tool calls, 51\% of cost) with limited validation (11\%--14\%). 
\textbf{Gemini CLI} shows a more balanced pattern across exploration (26\%--28\%), generation (21\%--41\%), and validation (18\%--35\%), and produces most outputs during generation (68\% of output tokens), indicating a more direct context-to-response flow.

Across agents, planning remains consistently low (<10\%), despite the open-ended nature of the task. 
Exploration and validation dominate both interaction and cost, suggesting that performance in agentic code review depends more on managing repository navigation and verification than on explicit planning.

\textbf{RQ2: How do agents operationalize command line tools in repository-grounded review?}
Tool usage patterns in Table~\ref{table:tool-use} highlight how agents operationalize review trajectories. 
\textbf{Rovo Dev} relies heavily on general-purpose shell commands across exploration and validation, indicating low-level navigation (e.g., \texttt{cd}, \texttt{grep}, \texttt{git}) and repeated context retrieval. 
\textbf{Claude Code} uses a more balanced mix of shell commands and structured tools (e.g., file reads and searches), reflecting a hybrid execution strategy. 
In contrast, \textbf{Gemini CLI} concentrates on a smaller set of specialized tools, mainly file reading and limited \texttt{git} operations, with low tool diversity.

\input{tables/top-tool-use}

In general, exploration is driven by file inspection and search primitives, generation by file-based comment writing, and validation by shell-based verification (e.g., \texttt{git}, \texttt{grep}). 
Overall, reliance on general-purpose shell commands—especially in Rovo Dev and Claude Code—suggests that agents often default to low-level execution rather than structured tooling, contributing to higher exploration overhead and less organized review trajectories.

\textbf{RQ3: How does computational cost incur during agentic code review?}
Trajectory analysis reveals clear efficiency trade-offs across agents. 
\textbf{Rovo Dev} produces the longest trajectories, averaging 7.4 validation and 5.5 exploration steps, and incurs the highest cost (\$0.75 validation, \$0.44 exploration) with over 10 validation tool calls per task. 
\textbf{Claude Code} exhibits similarly high exploration (7.5 steps) but substantially lower cost (\$0.10), indicating more efficient context retrieval despite comparable steps. 
In contrast, \textbf{Gemini CLI} operates with short trajectories ($\le$1 step per phase) and minimal cost across phases ($\le$\$0.007), reflecting a lightweight interaction.
In summary, exploration and validation dominate computational expenditure, highlighting that they are the primary cost drivers, while planning remains negligible (<1 step and near-zero cost).

\begin{figure}[t]
    \centering
    \includegraphics[width=0.9\linewidth]{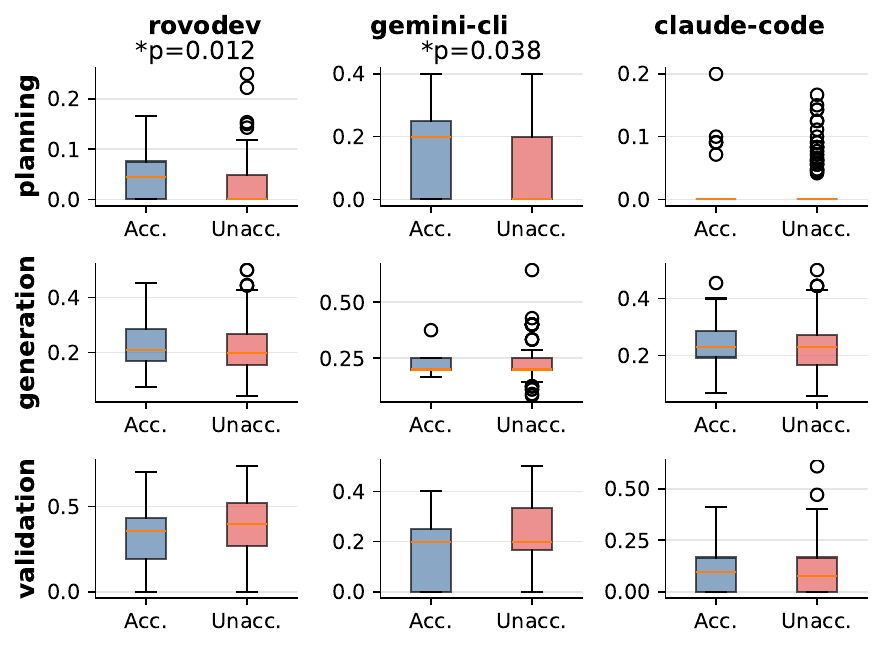}
    \caption{Accomplished vs. unaccomplished trajectory effort by phase and agent; significant planning differences appear for Rovo Dev and Gemini CLI ($p<0.05$).}
    \label{fig:accomplish-group-boxplot}
    \vspace{-6mm}
\end{figure}

\textbf{RQ4: How do successful and unsuccessful review trajectories differ?}
We compare trajectories from \emph{accomplished} and \emph{unaccomplished} reviews (as defined in Section~\ref{sec:trajectory_metric}), excluding partially accomplished cases to sharpen the contrast.

Figure~\ref{fig:accomplish-group-boxplot} shows that successful trajectories allocate significantly more effort to planning in \textbf{Rovo Dev} and \textbf{Gemini CLI} ($p<0.05$, MWU-Test), suggesting that explicit deliberation is associated with better outcomes. 
Generation effort is also slightly higher in successful cases, but not statistically significant. 
In contrast, unsuccessful trajectories show higher validation effort, which often corresponds to re-checking of context (e.g., reopening diffs and relevant files) without improving results. 
Overall, successful trajectories are more strongly associated with planning rather than increased validation.

\section{Implications and Conclusion}

\subsection{Implications}
Our findings suggest the following implications for agentic code review and its evaluation.
First, \textbf{localization is the bottleneck.}
Improving repository navigation likely yields greater gains than scaling model capability.
Second, \textbf{planning is the strongest predictor of success, yet the least invested phase.}
Planning consumes less than 10\% of effort despite being the only phase significantly associated with review success, suggesting agents are over-exploring and under-planning.
Third, \textbf{excessive validation signals weak initial grounding.}
Unsuccessful reviews exhibit heavier validation effort, yet repeated re-checking does not recover from poor planning.
\rev{Fourth, \textbf{cost does not imply performance.} 
The substantial cost gap between agents achieving comparable results highlights the need for cost-normalized evaluation metrics in future work.}

\subsection{Threats to Validity}

Manual verification and LLM-as-a-Judge evaluation may introduce noise, although we use an evaluated judge~\cite{tantithamthavorn2026hallujudgere}. 
Generated comments not aligned with ground truth should not be treated as incorrect, as agents may surface unforeseen issues. 
Trajectory phase labels rely on keyword-based heuristics, which may not fully capture latent intent \rev{(e.g., the same text may serve different purposes)}, though they are manually validated. 
Finally, findings may not generalize to future agents and systems; we release a replication package to support verification and future work.

\subsection{Conclusion}



We present the first trajectory-level empirical study of agentic code review in terminal environments, finding that localization is the key differentiator of agentic over static review, and that planning---despite being the least invested phase---is most strongly associated with review success.


\section*{Data Availability Statement}
The AgenticCR-Verified dataset, analysis scripts, and experimental results are available at https://doi.org/10.5281/zenodo.20131996.

\bibliography{references,sample-base}

\end{document}

%% file: tables/overall-performance.tex
\begin{table}[t]
\centering
\scriptsize
\caption{Terminal-based code review performance. Overall performance denotes the percentage of generated comments satisfying both localization and human-alignment criteria.}
\begin{tabularx}{\linewidth}{
    >{\hsize=1.3\hsize}X |
    >{\hsize=0.7\hsize}Y
    >{\hsize=0.9\hsize}Y
    >{\hsize=1.2\hsize}Y
    >{\hsize=0.9\hsize}Y
}

\toprule
\textbf{Code Reviewer} & \textbf{\# Comments} & \textbf{\% Localized Comments} & \textbf{\% Human Align. Comments} & \textbf{Overall Performance} \\
\midrule
\textbf{Rovo Dev 0.13.61}        & 860  & 20.93\% & 15.10\% & \textbf{5.81\%} \\
\textbf{Gemini CLI 0.32}    & 392  & 12.24\% & 13.28\% & 2.30\% \\
\textbf{Claude Code 2.1.72}    & 947  & \textbf{22.39\%} & 14.37\% & 5.60\% \\
\textbf{CodeRabbit 0.3.7}     & 803  & 7.22\%  & 8.54\%  & 1.37\% \\
\midrule
\textbf{GPT-5.2}        & 3180 & 7.86\%  & \textbf{21.44\%} & 2.61\% \\
\textbf{GPT-5.4}        & 2305 & 9.15\%  & 17.39\% & 2.43\% \\
\textbf{Gemini 2.5 Pro} & 3166 & 7.96\%  & 18.71\% & 2.37\% \\
\textbf{Claude Sonnet 4.5}     & 9060 & 7.70\%  & 12.68\% & 1.72\% \\
\textbf{Claude Sonnet 4.6}     & 4823 & 7.26\%  & 14.65\% & 1.87\% \\

\bottomrule
\end{tabularx}
\vspace{-7mm}
\label{tab:overall-performance}
\end{table}

%% file: tables/top-tool-use.tex
\begin{table}[t]
\centering
\scriptsize
\caption{Top frequently used tools (with representative commands) per phase and agent.}
\begin{tabularx}{\linewidth}{
    >{\hsize=0.7\hsize}X |
    >{\hsize=1.1\hsize}Y
    >{\hsize=1.1\hsize}Y
    >{\hsize=1.1\hsize}Y
}

\toprule
\textbf{Phase} & \textbf{Rovo Dev} & \textbf{Gemini CLI} & \textbf{Claude Code} \\
\midrule
Exploration 
& Shell (cd, git, grep), File Ops, Search 
& File Read, Shell (git), Search 
& Shell (git, grep, cat), File Read, Search \\
\hline
Planning 
& Shell (git, cd), File Ops 
& Shell (git) 
& Shell (git), Task Planning \\
\hline
Generation 
& File Write, Shell (cd, cat, git) 
& File Write 
& File Write, Shell (git) \\
\hline
Validation 
& Shell (cd, git, grep), Search, File Ops 
& Shell (git) 
& Shell (grep, git), Execution \\

\bottomrule
\end{tabularx}
\vspace{-5mm}
\label{table:tool-use}
\end{table}